\documentclass[namedreferences]{SolarPhysics}

\usepackage[optionalrh,natbib]{spr-sola-addons}
\usepackage{graphicx}
\usepackage{color}                       
\usepackage{url}    
\usepackage{a4} 
\usepackage{setspace}

\doublespace

\renewcommand{\vec}{\mathbf}
\renewcommand{\sun}{\astrosun}

\newcommand{\ut}[1]{\hat{\mathbf{#1}}}
\newcommand{\BE}{\begin{equation}}
\newcommand{\EE}{\end{equation}}

\begin{document}
\begin{article}

\begin{opening}

\title{Nonlinear force-free modeling of the corona in spherical coordinates}
\author{S.A. ~\surname{Gilchrist}$^{1}$\sep
        M.S. ~\surname{Wheatland}$^{1}$}

\institute{$^{1}$ Sydney Institute for Astronomy, School of Physics, The University of Sydney, NSW 2006, Australia }
            
\begin{abstract}
We present a code for solving the nonlinear force-free equations in 
spherical polar geometry, with the motivation of modeling the magnetic
field in the corona. The code is an implementation of the Grad-Rubin
method. Our method is applicable to a spherical domain of arbitrary angular
size. The implementation is based on a global spectral representation for the 
magnetic field which makes no explicit assumptions about the form of the magnetic
field at the transverse boundaries of the domain. We apply the code to 
a bipolar test case with analytic boundary conditions, and we 
demonstrate the convergence of the Grad-Rubin method, and the 
self-consistency of the resulting numerical solution. 
\end{abstract}


\end{opening}


%
%
\section{Introduction}

Basic properties of the solar coronal magnetic field such as its strength,
direction, and three-dimensional structure, cannot be presently determined
by observation, and this motivates modeling of the coronal magnetic field.
In the corona above active regions the pressure and gravity forces 
are too small to balance the magnetic (Lorentz) force \citep{1995ApJ...439..474M,2001SoPh..203...71G}
so it is common to model the coronal magnetic field as force-free, {\it i.e.} a magnetic 
field where the Lorentz force is identically zero, and electric currents flow along magnetic field lines \citep{1994ppit.book.....S}. 
Force-free models of the corona were reviewed by 
\citet{1989SSRv...51...11S}, and \citet{livrev2012}. 

A force-free magnetic field satisfies \citep{1994ppit.book.....S} 
\BE
  \nabla \times \vec B = \alpha(\vec r) \vec B,
  \label{ff1}
\EE
and 
\BE
  \nabla \cdot \vec B = 0, 
  \label{ff2}
\EE 
where the scalar field $\alpha(\vec r)$ is 
related to the electric current density in the volume $\vec J$ by
\BE
  \vec J = \alpha(\vec r) \vec B/\mu_0.
  \label{ff3}
\EE
Equations (\ref{ff1}) and (\ref{ff2}) require boundary conditions to determine a solution, and
this defines the force-free boundary value problem. The correct 
boundary conditions for a well-posed formulation of the problem were 
outlined by \citet{gr}. In the Grad-Rubin formulation the boundary
conditions are the normal component of the magnetic field, $B_n$,  
and the value of $\alpha$ over one polarity of the field in the boundary, {\it i.e.}
values of $\alpha$ are specified over points in the boundary where $B_n<0$,
or where $B_n>0$. For modeling the coronal magnetic field the boundary conditions are assumed
at the photosphere and the model is solved in the coronal volume. In 
theoretical studies the boundary conditions usually have an analytic form ({\it e.g.} 
\citealt{2009A&A...497L..17R,2012SoPh..277..131R}), and in observationally-based
studies the boundary conditions are typically derived from spectro-polarimetric observations
of the photospheric magnetic field ({\it e.g.} \citealt{2008ApJ...675.1637S,2009ApJ...696.1780D},
and also see references in \citealt{livrev2012}).

The force-free boundary value problem is nonlinear in the general case
where $\alpha$ is a function of position. When $\alpha$ is constant
the equations are linear and closed form analytic solutions can be found 
\citep{1972SoPh...25..127N,1981A&A...100..197A}. However, the linear model 
is unphysical in that the solutions in general have unbounded energy in an 
unbounded space \citep{1981A&A...100..197A}. For the nonlinear problem, 
analytic solutions can be found using the generating function method \citep{1994ppit.book.....S} for
particular symmetries, for example for rotational symmetry \citep{1990ApJ...352..343L}.
The general nonlinear problem has no known analytic solution and must be 
treated numerically. For this purpose a number of methods have been developed.
These methods differ in their formulation of the boundary value problem and their choice 
of solution method (for reviews see \citet{2008JGRA..11303S02W} and 
 \citet{livrev2012}).

Most of the force-free methods in use solve the force-free equations
in Cartesian geometry, with the corona corresponding to the half-space
$z>0$ and the photosphere represented by the $z=0$ plane \citep{1981SoPh...69..343S,
1990ApJ...362..698W,2000ApJ...540.1150W,2004SoPh..219...87W,
2007SoPh..245..251W,2005A&A...433..335V}. This introduces two
problems into the modeling. The first problem is that the Cartesian 
approximation, which assumes that the curvature of the Sun in negligible, 
becomes inaccurate when considering large regions on the Sun \citep{1990SoPh..126...21G}.  
Full-disk spectro-polarimetric observations of the photospheric magnetic
field are now available from the {\it Helioseismic and Magnetic Imager} 
(HMI: \citealt{2012SoPh..275..229S}) aboard the {\it Solar Dynamics Observatory} 
(SDO:  \citealt{2012SoPh..275....3P}). Coronal field modeling based on
these data must use spherical coordinates. The second problem concerns the assumption of boundary conditions on 
the transverse boundaries of the Cartesian domain. In practice the
infinite half-space is replaced by a finite numerical domain, meaning
boundary conditions are required on the top and side boundaries of the 
volume in addition to the $z=0$ plane. In general, ad hoc boundary conditions are 
used, such as assuming periodicity, or assuming that no
magnetic flux leaves the top or side boundaries. These boundary conditions are
artificial and do not necessarily represent physical conditions in the corona. 
Spherical modeling avoids this problem, because a spherical domain can encompass the entire corona  
with no transverse boundaries, in which case ad hoc boundary 
conditions are not required. 

Modeling the entire corona avoids the need for boundary conditions
apart from those at the photosphere, but introduces other difficulties. The description of the polar field
presents difficulties both observationally and numerically. 
The observational difficulties are two-fold. Firstly, due to the Sun's tilt, 
only one pole is observed from Earth at a time. 
Secondly, spectro-polarimetrically derived magnetic field values may be
inaccurate near the poles where there may be significant unresolved 
mixed-polarity magnetic flux. This can lead to partial cancellation of 
the polarization signal at each pixel. Quiet-Sun regions also contain
mixed polarity flux, but the problem is likely to be worse 
close to the poles due to line-of-sight effects. For these
reasons the polar field is usually interpolated from observations
at lower latitudes ({\it e.g.} \citealt{2011SoPh..270....9S}). The co-ordinate
singularities in the spherical polar system which occur at the poles 
also pose a problem for numerical methods. For finite difference 
methods the coordinate poles require special treatment, and with spectral 
methods specific grids are required to avoid problems \citep{boyd_spectral}.

In this paper we outline an implementation of the Grad-Rubin method
for solving the nonlinear force-free model in spherical geometry. 
Our method is applicable both to the entire Sun and to regions with restricted 
angular extent. Other methods for solving the force-free model 
in spherical geometry have been developed. \citet{2007SoPh..240..227W}
presented a generalization of the optimization method, and 
\citet{2013A&A...553A..43A} presented a finite-difference implementation of the Grad-Rubin method. 
Our method differs from both of these and features a new spectral method for 
computing the magnetic field based on an expansion of the field in terms
of global basis functions. This solution can be applied to a spherical
region of the corona of arbitrary angular size, in which case explicit 
assumptions about the magnetic field on the transverse
boundaries of the spherical region are not required, even for a spherical
wedge with restricted extent. However, for the latter case we impose additional constraints on 
the boundary conditions at the photosphere. Specifically, we assume that 
the radial component of the field and the electric current density vanish at the photosphere outside the wedge.
We present the application of our method to 
a simple test case with analytic boundary conditions to demonstrate the 
convergence of the method and the self-consistency of the solution. 

This paper is structured as follows.
In Section \ref{s_theory} we outline the specific form of the force-free
boundary value problem that we solve, and we outline 
the Grad-Rubin iteration method. In Section \ref{sec3} we present the 
details of our implementation of the Grad-Rubin method, including the spectral 
solution for the magnetic field that is used. In Section \ref{sec4} we describe 
the test case that we use and show the results of applying the code to the test case. 
Finally, in Section \ref{sec5} we present a discussion of the results
and a conclusion.

%
%
\section{Theory}
\label{s_theory}

In this section we outline the two boundary value problems that we solve, 
and give a brief description of the Grad-Rubin method. We consider 
solutions of the nonlinear force-free Equations (\ref{ff1}) and (\ref{ff2}) in two domains. The first 
domain is the entire corona, and the second is a spherical wedge of 
limited angular extent.  

%
%

\subsection{Boundary value problems}
\label{ss_bvp}

We first consider the domain $\Omega_{\rm global}$ which is defined
as the set of points with spherical polar coordinates
\BE
  \Omega_{\rm global} = \{(r,\theta,\phi) \;| \;r\in[R_{\sun},\infty), \theta\in(0,\pi),\phi\in [0,2\pi) \},
  \label{dom_global} 
\EE
where $\theta$ is the polar angle,
$\phi$ is the azimuthal angle, $r$ is the radius, and $R_{\sun}$ is the 
radius of the Sun. We refer to this domain as global because it 
covers a complete $4\pi$ steradians, and its lower boundary is the
entire photosphere.

It is necessary to specify boundary conditions on the force-free
equations at the photosphere. Following the prescription of \citet{gr},
the appropriate boundary conditions are the radial component of $\vec B$, 

\BE
  \vec B \cdot \ut{r} |_{r=R_{\sun}} = B_n(\theta,\phi),
\EE
and the force-free parameter 
\BE
  \alpha |_{r=R_{\sun}} = \alpha_0(\theta,\phi) 
\EE
over one polarity of $B_n$, {\it i.e.} values of $\alpha_0$ are specified either where
$B_n<0$ or where $B_n>0$. It is also assumed that the magnetic
field vanishes for large $r$, {\it i.e.} 
\BE
  \lim_{r \rightarrow \infty} \vec B = 0. 
  \label{asym}
\EE 
This asymptotic boundary condition matches that used by some 
Cartesian codes \citep{2007SoPh..245..251W}.

In some cases $B_n$ and $\alpha_0$ may only be non-zero over 
a small range of $\theta$ and $\phi$. In this situation it is 
unnecessary to use a global domain. A more appropriate 
choice is 
\BE
  \Omega_{\rm wedge} = \{(r,\theta,\phi) \; | \; r\in[R_{\sun},\infty), \theta\in[\theta_{\rm min},\theta_{\rm max}],
  \phi\in [\phi_{\rm min},\phi_{\rm max}] \},
  \label{dom_wedge} 
\EE
{\it i.e.} a domain external to a sphere but restricted in angular extent. We refer to this as 
a spherical wedge. In principle, it is necessary to prescribe boundary conditions
at the transverse boundaries of this domain, but in Section
\ref{sec3} we explain how to obtain solutions for which this is unnecessary. 
This approach assumes that $B_n$ and $\alpha_0$ are zero 
everywhere outside the domain $\Omega_{\rm wedge}$.

Here we are using the Grad-Rubin boundary conditions which assume $\alpha_0$
over a single polarity of $B_n$, but observational data provides $\alpha_0$
over both \citep{2004ASSL..307.....L}. This means two possible solutions can in principle be found for a given
data set, corresponding to the two choices of polarity. If the data
are consistent with the force-free model, then the two solutions will
be the same. However, in practice it is found that the two solutions
differ significantly ({\it e.g.} \citealt{2008ApJ...675.1637S}). This may
be attributed to the departure of the photospheric field from the
force-free state due to significant pressure and gravity forces \citep{1995ApJ...439..474M}. 
\citet{2009ApJ...700L..88W} presented a method, based on 
an implementation of the Grad-Rubin method in Cartesian coordinates, for using 
the data from both polarities to construct a single self-consistent 
force-free solution. Here we present only the basic Grad-Rubin method,
but in principle the \citet{2009ApJ...700L..88W} self-consistency 
procedure may be applied here also.

%
%

\subsection{Grad-Rubin iteration}
\label{ss_gri}

The Grad-Rubin method is an iterative method for solving the nonlinear
force-free equations \citep{gr}. The method has previously been implemented in 
Cartesian coordinates \citep{1981SoPh...69..343S,1999A&A...350.1051A,2007SoPh..245..251W}
and in spherical coordinates \citep{2013A&A...553A..43A}. The method replaces the nonlinear 
Equations (\ref{ff1}) and (\ref{ff2}) with a pair of linear equations which are solved 
repeatedly in a sequence of iterations. We denote a quantity after $n$ Grad-Rubin iterations using
a superscript in square brackets, {\it e.g.} $\vec B^{[n]}$. One Grad-Rubin
iteration may be written  

\BE
  \nabla \alpha ^{[n+1]} \cdot \vec B^{[n]} = 0,
  \label{gr1}
\EE
and 
\BE
  \nabla \times \vec B^{[n+1]} = \alpha^{[n+1]} \vec B^{[n]}.
  \label{gr2}
\EE
Equation (\ref{gr1}) updates the force-free parameter $\alpha$ in the volume subject to boundary 
conditions on $\alpha$. Equation (\ref{gr2}) updates the magnetic
field in the volume using the new $\alpha$ values together with the 
magnetic field from the previous iteration subject to the boundary
conditions on the normal component of the field. Equations (\ref{gr1})
and (\ref{gr2}) are repeatedly solved until the magnetic field $\vec B^{[n]}$
and the force-free parameter $\alpha^{[n]}$ converge at all points
in the volume. The iteration is initiated using a potential field $\vec B_0$ 
constructed from the boundary conditions on $B_n$. We present the details 
of our method for solving Equations (\ref{gr1}) and (\ref{gr2}) numerically 
in two spherical domains in Section \ref{sec3}.

%
%
\section{Numerical implementation} 
\label{sec3}

In this section we outline our implementation of the Grad-Rubin method
in code. The Grad-Rubin method requires an initial potential field,
a method for solving Equation (\ref{gr1}) to update $\alpha$, and
a method for solving Equation (\ref{gr2}) to update the magnetic field.

The numerical grid used is a spherical polar grid with $N_r$ points
in the $r$ direction, $N_{\theta}$ points in the $\theta$ direction, 
and $N_{\phi}$ points in the $\phi$ direction. The $\phi$ and $r$ grids are uniformly spaced. The $\theta$ grid is either
a Gauss-Legendre grid or is uniform \citep{2007nrfa.book.....P}. The Gauss-Legendre grid is 
required to accurately represent the solution near the poles and is 
only necessary for constructing solutions in the global domain. The
grid is finite in the radial direction, and has a maximum $r$ value
which we call $R_{\rm max}$.

%
%

\subsection{Spectral solution for the potential field} 
\label{ss_ssftpf}

An initial potential field is calculated as a starting point for the Grad-Rubin iteration. 
We use a spherical harmonic solution for the potential field. The spherical
harmonics are global basis functions, meaning they are orthogonal
over the domain $\Omega_{\rm global}$. It can be 
shown that, in terms of spherical harmonics $Y_{lm}(\theta,\phi)$, the 
components of the potential field satisfying the boundary condition
Equation (\ref{asym}) at infinity are \citep{1969SoPh....9..131A}

\BE
  B_r = \sum \limits_{l=0}^{\infty} \sum \limits_{m=-l}^{l} a_{lm} 
  \left (\frac{R_{\sun}}{r} \right )^{(l+2)} Y_{lm}(\theta,\phi),
  \label{br_pot}
\EE
\BE
  B_{\theta} = \sum \limits_{l=0}^{\infty} \sum \limits_{m=-l}^{l}  
  -\frac{a_{lm} }{l+1} 
  \left (\frac{R_{\sun}}{r} \right )^{(l+2)} 
  \frac{\partial Y_{lm}(\theta,\phi)}{\partial \theta}
  \label{bt_pot}
\EE
and
\BE
  B_{\phi} = \sum \limits_{l=0}^{\infty} \sum \limits_{m=-l}^{l}  
  -\frac{ima_{lm} }{l+1} 
  \left (\frac{R_{\sun}}{r} \right )^{(l+2)} 
  \frac{Y_{lm}(\theta,\phi)}{\sin\theta},
  \label{bp_pot}
\EE
where the coefficients $a_{lm}$ are given by 
\BE
  a_{lm} = \int \limits_0 ^{2\pi} \int \limits_0^{\pi} B_n(\theta,\phi)
  Y^{*}_{lm}(\theta,\phi) \sin\theta d\theta d\phi,
  \label{co1}
\EE
and where $i^2=-1$. These equations are complex valued and the physical magnetic
field is the real part. Equations (\ref{br_pot})-(\ref{co1}) can be obtained from the 
well-known potential source-surface solution \citep{1969SoPh....9..131A}
by considering that solution in the limit where the source surface
is located at infinity. 

In practice the series must be truncated after a finite number of 
terms. We truncate the series at a finite $l$ value which we call
$L$, {\it i.e.} we perform the summation over all the spherical harmonics
with $l \le L$ and $|m| \le l$. This approach results in a truncation error 
in $\vec B$ which is position independent \citep{boyd_spectral}.  
The series is a Fourier series in $\phi$, so it is natural 
to choose $L$ to correspond to the Nyquist frequency \citep{boyd_spectral}:
\BE
L = \frac{\pi}{\Delta \phi},
\label{niq}
\EE
where $\Delta \phi$ is the uniform spacing of points in 
$\phi$. In practice the right hand side of Equation (\ref{niq}) is  
rounded to the nearest integer. 

Equations (\ref{br_pot})-(\ref{co1}) provide the magnetic field 
at all points in the global domain $\Omega_{\rm global}$. The solution
in the restricted domain $\Omega_{\rm wedge}$ can be found by 
evaluating the global solution only at points contained in $\Omega_{\rm wedge}$.  
This allows the solution in $\Omega_{\rm wedge}$ to be found without 
assuming specific boundary conditions on the transverse boundaries 
of $\Omega_{\rm wedge}$. Since we are assume that $B_n$ is zero 
outside $\Omega_{\rm wedge}$, the integral in Equation (\ref{co1})
need only be computed over the restricted domain. 

\citet{2011ApJ...732..102T} report the non-convergence
of the spectral series given by Equations (\ref{bt_pot})-(\ref{bp_pot}).
The non-convergence results in erroneous magnetic field values 
(particularly near the poles) for large $L$. The problem occurs because 
the numerical grid used by \citet{2011ApJ...732..102T} is not sufficiently
dense near the poles to accurately represent the rapid variation of the spherical harmonics. 
For calculations in $\Omega_{\rm global}$ 
we use a Gauss-Legendre grid which accurately represents the spherical
harmonics near the poles \citep{boyd_spectral}. For calculations in $\Omega_{\rm wedge}$, we use uniform  
grid in $\theta$, which is rotated such that the region of
interest is isolated from the poles, and does not encounter
this problem.

The Gibbs phenomenon (ringing produced in representing 
discontinuous changes) is a problem for all spectral methods \citep{boyd_spectral}. The problem
is significant when spectral potential field solutions are calculated from
observational data \citep{2011ApJ...732..102T}. Including more terms
in the series improves but does not eliminate the problem. 
It is important to note this particular caveat when applying
and interpreting results produced by spectral methods. It should  
be noted that finite difference methods also become inaccurate at locations
with large gradients in the field being represented. 

A parallel code is used to sum the spectral series. The 
coefficients $a_{lm}$ are calculated using Equation (\ref{co1}) 
and then Equations (\ref{br_pot})-(\ref{bp_pot}) are evaluated with
the sums performed using partial sums, {\it i.e.} each series is broken into a number
of sub-series, each of which is summed independently, and then the final 
result is obtained by adding the partial sums. The parallel implementation
is written for a distributed memory multiprocessor. The method 
uses a combination of the Message Passing Interface (MPI) \citep{mpiref} 
and OpenMP \citep{openmp} and is described in Appendix B. 

%
%

\subsection{Field line tracing solution for the current-update step}

To solve Equation (\ref{gr1}) we employ the field line tracing method 
which has been used in Grad-Rubin implementations in Cartesian 
coordinates \citep{1999A&A...350.1051A,2007SoPh..245..251W}, and in
spherical coordinates \citep{2013A&A...553A..43A}. According to Equation 
(\ref{gr1}), $\alpha^{[n+1]}$ is constant along magnetic
field lines. The field line tracing method determines $\alpha^{[n+1]}$
in the volume by tracing the field line threading each grid point until it 
crosses the lower boundary, and the value of $\alpha_0$ at the crossing point 
in the boundary is assigned to the grid point. The field line is traced in
the forwards direction if boundary values for $\alpha_0$ are 
chosen where $B_n<0$, and is traced in the backwards direction
if boundary values for $\alpha_0$ are chosen where $B_n>0$. 
Points in the volume connected to field lines which leave the domain through the outer 
boundary $r=R_{\rm max}$ are assigned $\alpha^{[n+1]}=0$ at the point
in the volume. In addition, points in the volume threaded by
field lines which leave the transverse boundaries of $\Omega_{\rm wedge}$
are assigned $\alpha^{[n+1]}=0$ at the point in the volume. The tracing is performed using fourth-order
Runge-Kutta integration  \citep{2007nrfa.book.....P}, and trilinear
interpolation is used to determine $\vec B^{[n]}$ at points along 
the field line not coinciding with a grid point. 

%
%

\subsection{Spectral solution to Ampere's law for the field-update step}

To solve Equation (\ref{gr2}) we use a spectral solution. 
The magnetic field is decomposed into the sum of a 
potential field and a non-potential field, {\it i.e.} 
\BE
  \vec B^{[n+1]} = \vec B_0 + \vec B^{[n+1]}_c,
\EE
where $\vec B^{[n+1]}_c$ satisfies 
\BE
  \nabla \times \vec B^{[n+1]}_c=\vec J^{[n+1]}, 
  \label{ampere_gr}
\EE
with 
\BE
  \vec J^{[n+1]}=\alpha^{[n+1]}\vec B^{[n]}/\mu_0, 
  \label{jdef}
\EE
and where $\vec B_0$ is the potential field matching the boundary conditions on $B_n$ calculated
using the method of Section \ref{ss_ssftpf}. It is only necessary to update 
$\vec B^{[n+1]}_c$ at each iteration as $\vec B_0$ does not change. Also,
since $\vec B_0$ satisfies the boundary conditions on the normal 
component of the field at $r=R_{\sun}$ it follows
that 
\BE
  \vec B^{[n+1]}_c|_{r=R_{\sun}} = 0,
  \label{np_bc1}
\EE
and from Equation (\ref{asym}) we require
\BE
  \lim_{r \rightarrow \infty} \vec B^{[n+1]}_c = 0.
  \label{np_bc2}
\EE
Equations (\ref{np_bc1}) and (\ref{np_bc2}) define the boundary
conditions on $\vec B^{[n+1]}_c$. 

We use a spectral solution to Equation (\ref{ampere_gr}) which is analogous to the 
spherical harmonic solution for the potential field. We express
$\vec B^{[n+1]}_c$ as a series using the vector spherical harmonics \citep{1953mtp..book.....M},
{\it i.e.}
\BE
\vec B^{[n+1]}_c = \sum \limits_{l=0}^{\infty} \sum \limits_{m=-l}^{l}
  B^{(1)}_{lm}(r) \vec Y_{lm} + B^{(2)}_{lm}(r) \vec \Psi_{lm} + 
  B^{(3)}_{lm}(r) \vec \Phi_{lm}, 
\EE
where $\vec Y_{lm}$, $\vec \Psi_{lm}$, and $\vec \Phi_{lm}$ are 
the complete set of orthogonal vector basis functions defined by
\BE
  \vec Y_{lm} = Y_{lm}{\hat \vec r},
\EE 
\BE
  \vec \Psi_{lm} = \frac{r \nabla Y_{lm}}{\sqrt{l(l+1)}},
\EE
and
\BE
  \vec \Phi_{lm} = \frac{\vec r \times \nabla Y_{lm}}{\sqrt{l(l+1)}}.
\EE
These functions are mutually perpendicular, {\it i.e.}
\BE
  \vec Y_{lm} \cdot \vec \Psi_{lm} = \vec Y_{lm} \cdot \vec \Phi_{lm} = 
  \vec \Psi_{lm} \cdot \vec \Phi_{lm} = 0,
\EE
and orthonormal, {\it e.g.} 
\BE
  \int \limits_0^{\pi} \int \limits_0^{2\pi} \vec Y_{lm}(\theta,\phi)\cdot \vec 
  Y_{l'm'}^{*}(\theta,\phi) d\theta d\phi= \delta_{l l'}\delta_{m m'},
\EE
where $\delta_{lm}$ is the Kronecker delta\footnote{A similar integral relation
applies for $\vec \Psi_{lm}$ and $\vec \Phi_{lm}$.}. The vector spherical
harmonics have previously been applied to magnetostatic problems
({\it e.g.} \citealt{vshmag,denquart}), but have not been used in this 
context. 

The spectral coefficients $B^{(i)}_{lm}$ with $i=1,2,3$ are determined by the distribution of currents 
in the volume and by the boundary conditions. We show in Appendix A that for the problem
at hand the spectral coefficients for the magnetic field are
\BE
  B^{(1)}_{lm} = \frac{\sqrt{l(l+1)}}{r} \left [ -R_{\sun}^l \left ( \frac{R_0}{r} \right )^{l+1} I_0 + I_2(r) + I_3(r)
  \right ],
\label{b1_comp}
\EE
\BE
  B^{(2)}_{lm} = \frac{1}{r} \left [ R_{\sun}^l \left (\frac{R_{\sun}}{r} \right )^{l+1}I_0
  - lI_1(r) + (l+1)I_2(r)  \right ],
\label{b2_comp}
\EE
and 
\BE
B^{(3)}_{lm} =  \frac{r J^{(1)}_{lm}}{\sqrt{l(l+1)}},
\label{b3_comp}
\EE
where 
\BE
  I_0 = \frac{\mu_0}{2l+1} \int^{\infty}_{R_{\sun}} s^{1-l} J^{(3)}_{lm}(s) ds,
  \label{int0}
\EE
\BE
  I_1(r) = \frac{\mu_0}{2l+1} \int^{r}_{R_{\sun}} s \left ( \frac{s}{r} \right )^{l+1}J^{(3)}_{lm}(s)ds,
  \label{int1}
\EE
and 
\BE
  I_2(r) = \frac{\mu_0}{2l+1} \int^{\infty}_{r} s \left ( \frac{r}{s} \right )^l J^{(3)}_{lm}(s)ds.
  \label{int2}
\EE
The coefficients $J_{lm}^{(i)}$ are the spectral coefficients of the 
current distribution defined by 
\BE
  J_{lm}^{(1)}(r) = \int \vec J^{[n+1]} \cdot \vec Y^{*}_{lm} d\Omega,
  \label{jj1}
\EE
\BE
  J_{lm}^{(2)}(r) = \int \vec J^{[n+1]} \cdot \vec \Psi^{*}_{lm} d\Omega,
  \label{jj2}
\EE
and
\BE
  J_{lm}^{(3)}(r) = \int \vec J^{[n+1]} \cdot \vec \Phi^{*}_{lm}d\Omega,
  \label{jj3}
\EE
where $\vec J^{[n+1]}$ is the volume current density defined by Equation (\ref{jdef}).
The spectral solution is computed in three steps: i) 
$J^{(i)}_{lm}$ is computed from $\vec J^{[n+1]}$ using Equations (\ref{jj1})-(\ref{jj3});
ii) spectral coefficients for the magnetic field are computed from 
Equations (\ref{b1_comp})-(\ref{b3_comp}); iii) the spectral series
is summed to a maximum order $L$. A parallel summation method is used,
as described in Appendix B (the method is described for the potential
field calculation, but the same approach is also used for the non-potential
component of the field). 

The integrals $I_0$, $I_1$, and $I_2$ [Equations (\ref{int0}), (\ref{int1}), 
and (\ref{int2})] are evaluated using the trapezoidal rule. The integrals are expressed in 
a such a way as to avoid numerical overflow for large values of $l$. Although
the integrals in Equations (\ref{int0}) and (\ref{int2}) are written
as extending to infinite radius, in the numerical solution the maximum radius is
$R_{\rm max}$, and an error is introduced by this approximation if $\vec J^{[n+1]} \ne 0$ at 
$r=R_{\rm max}$. Therefore it is necessary to make $R_{\rm max}$ 
sufficiently large to encompass all the significant currents.

%
%
\section{Application to test cases} 
\label{sec4}

In this section we apply our code to two test cases with analytic 
boundary conditions, to demonstrate the method. We establish the 
convergence of the method and quantify the self-consistency of the solution. 

%
%

\subsection{Two bipolar test cases}
\label{sec_btc}

The first test case is a simple model with analytic boundary conditions, representing the field due 
to a bipolar active region with Gaussian sunspots calculated in
the global domain $\Omega_{\rm global}$. A small non-zero patch of $\alpha$ is included around 
one of the spots. The second test case has the same boundary conditions,
on a smaller spacial scale, and the field is calculated in a restricted 
domain $\Omega_{\rm wedge}$. For each test case we demonstrate the convergence of the Grad-Rubin
iteration, and we measure the self-consistency of the solution by 
verifying that the Lorentz force in the model corona is zero, {\it i.e.}  that the electric current
is parallel to the magnetic field.

For both test cases, the boundary conditions on the magnetic 
field are
\BE
  B_n(\theta,\phi) = B_{\rm s} \left ( e^{-s_1^2/\sigma^2} - e^{-s_2^2/\sigma^2} \right ),
  \label{test_case1}
\EE
where $B_{\rm s}$ is a scale constant which is chosen such that $\mbox{max}{(|B_n|)}=B_{\rm s}$, and $\sigma$ is a 
parameter which determines the size of the spots. The two functions
$s_1$ and $s_2$ are distances to the centers of each spot as measured
on the sphere, and may be
written 
\BE
  s_i(\theta,\phi) = R_{\sun} \tan^{-1} \left [ \frac{\sqrt{(\sin\theta\sin\Delta \lambda)^2 + 
  (\sin\theta_i\cos\theta-\cos\theta\sin\theta_i\cos \Delta \lambda)^2}}
  {\cos\theta_i\cos\theta+\sin\theta\sin\theta_i\cos\Delta\lambda} \right ],
\EE
where $\Delta \lambda = \phi_i - \phi$, and with $(\theta_1,\phi_1)$ being 
the coordinates of the center of the spot with positive polarity, and with $(\theta_2,\phi_2)$ being  
the coordinates of the center of the spot with negative
polarity. For the boundary conditions on $\alpha_0$ we consider an isolated spot with
a locally constant value of $\alpha$ around the center of the positive spot, {\it i.e.}  
\BE
  \alpha_0 = \left \{
  \begin{array}{l l}
    \Lambda & \quad  B_n \ge B_{\rm th} \\ 
    0      & \quad  0 < B_n < B_{\rm th}
  \end{array} \right. ,
\EE
where $\Lambda$ is a constant, and $B_{\rm th}$ is a
threshold value. Note that the boundary conditions
on $\alpha_0$ are only defined over one polarity of $B_n$ 
(as per the Grad-Rubin formulation). We have chosen to define 
$\alpha_0$ over the positive polarity of $B_n$, {\it i.e.}  where $B_n>0$. 

It is known that force-free fields can be unstable when
\BE
  \alpha_{\rm s} L_{\rm s} > 1 
\EE
\citep{1974SoPh...39..393M}, where $\alpha_{\rm s}$ is a characteristic value for $\alpha$, and $L_{\rm s}$ is the characteristic scale length. For 
the bipolar test case we take the scale length to be the distance 
along the photosphere between the two spots, and we take $\alpha_{\rm s}=\Lambda$.
For both test cases we choose these values such that 
\BE
  \alpha_{\rm s} L_{\rm s} = 1,   
\EE
corresponding to a current matching the approximate maximum.

%
%

\subsection{Measures of convergence } 

To measure the convergence of the iteration we employ two metrics. 
The first is the total magnetic energy at each iteration $n$:
\BE
  E_n = \frac{1}{2\mu_0} \int  |{\vec B^{[n]}} |^2 dV,
\EE
where the integral is over the entire computational volume. We 
expect the energy to converge to a fixed value with Grad-Rubin iteration. 
The second metric is the average absolute difference in the magnetic field
at successive Grad-Rubin iterations: 
\BE
  A_{\rm mean} = \langle |\vec B^{[n]}-\vec B^{[n-1]}| \rangle, 
\EE
where $\langle ... \rangle$ denotes the average over the computational
volume. We expect $A_{\rm mean}$ to approach zero with repeated 
iteration.

To measure the self consistency of the solution, we consider 
the maximum angle between $\vec J$ and $\vec B$. The angle is defined by 
\BE
  \theta_{\rm max} = {\rm max}\left [ \frac{|\vec J^{[n]} \times \vec B^{[n]}|}{|\vec B^{[n]}||\vec J^{[n]}|} \right ],
\EE
where the maximum is calculated over all grid points. 

We also examine the self-consistency qualitatively by drawing the 
field lines of $\vec B^{[n]}$ and streamlines of the current density 
$\vec J^{[n]}$. For an exact force-free solution we expect that these 
two sets of field lines will coincide corresponding to $\theta_{\rm max}=0$. This 
provides  quite a stringent test, because the error in the paths of
the field lines is the result of the local truncation error 
in the solution integrated along the paths, which is in general 
greater than the local error. 

%
%

\subsection{Test case one: the bipole in the global domain}
\label{tcone_bg}

The first test case described in Section \ref{sec_btc} considers
a bipolar active region covering a significant fraction of the photosphere.
This region provides a test for the method in the domain $\Omega_{\rm global}$.
The parameters used for this test are shown in Table \ref{t1}, and the dimensions of the grid
are summarized in Table \ref{t2}. We perform 40 Grad-Rubin iterations
starting from a potential field. The computation takes approximately 
15 minutes on a computer with an eight-core CPU.

\begin{figure}
  \centerline{\includegraphics[scale =0.75]{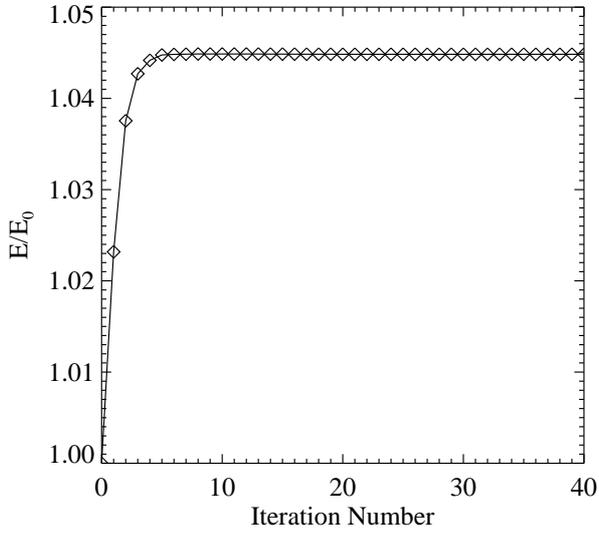}\includegraphics[scale =0.75]{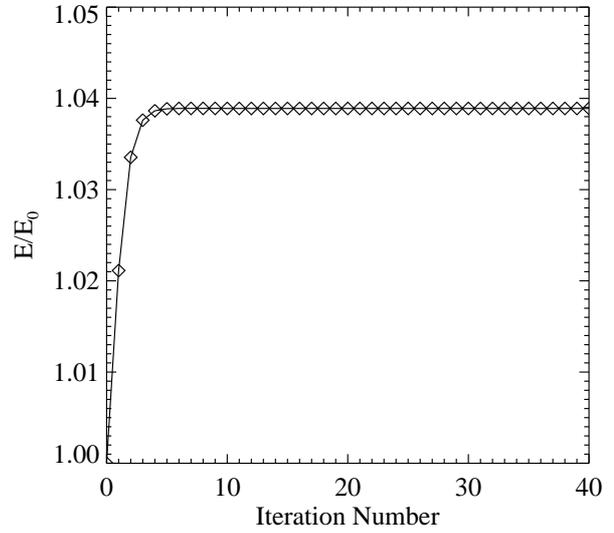}}
  \caption{Energy (in units of the energy of the potential field $E_0$)
           as a function of iteration number. The left panel shows $E/E_0$
           for the first test case in the global domain $\Omega_{\rm global}$, 
           and the right panel shows the same for the second test case, involving 
           a calculation in the domain $\Omega_{\rm wedge}$. In both cases the energy converges to 
           an approximately constant value after about six iterations.} 
\label{f1}
\end{figure}

The left panel of Figure \ref{f1} illustrates the
energy (in units of the energy of the potential field $E_0$) as a function of iteration number. 
The energy converges to an approximately constant value after about 
six Grad-Rubin iterations. The left panel of Figure \ref{f2} shows the
absolute average change in the field, $A_{\rm mean}$, as a function of iteration number. 
The scale on the $y$ axis is logarithmic. This figure shows that $A_{\rm mean}$
decreases exponentially before becoming roughly constant. The behavior
of the two metrics $E/E_0$ and $A_{\rm mean}$ establishes the convergence
of the Grad-Rubin procedure for this case.

\begin{figure}
  \centerline{\includegraphics[scale =0.5]{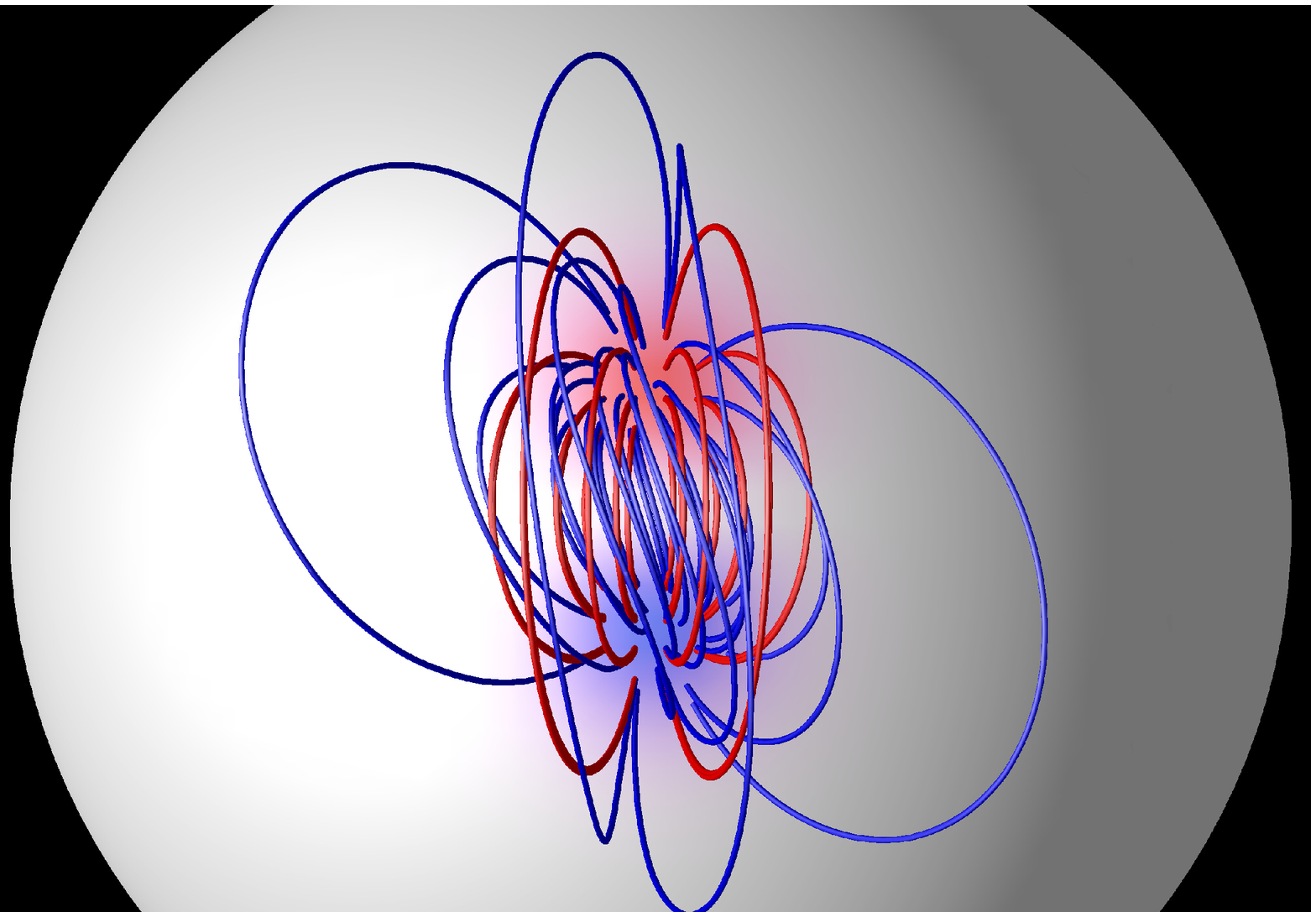}
              \includegraphics[scale =0.5]{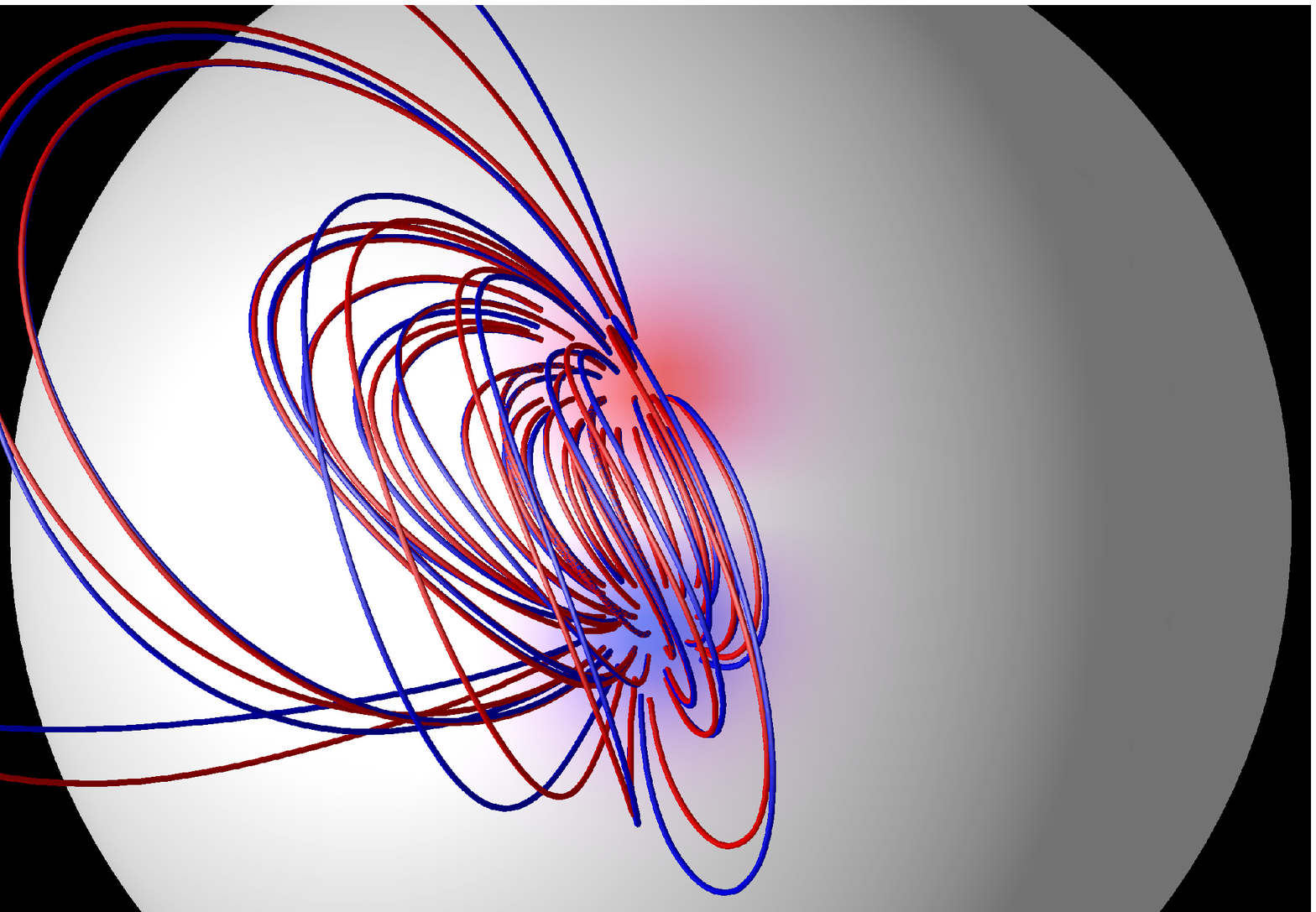}}
  \caption{Field lines (blue) and streamlines of current density (red)
           for test case one. The left panel is after the first 
           Grad-Rubin iteration. The right is after 40 Grad-Rubin iterations. In
           the left panel there is a clear difference between the two 
           sets of lines, and in the right the two sets of lines coincide, indicating 
           that a force-free solution has been found. The discrepancy between the two 
           sets of lines is due to the interpolation of field values between
           grid points necessary to draw the lines (see the discussion in Section \ref{tcone_bg}).} 
\label{f3}
\end{figure}

We also demonstrate the self-consistency of the numerical solution. 
Figure \ref{f3} shows the field lines of $\vec B$ (in blue) and 
the streamlines of $\vec J$ (in red). The left panel shows the results after
one Grad-Rubin iteration. In this panel there is a significant discrepancy
between the two sets of lines. The right panel shows the field 
lines after 40 Grad-Rubin iterations. In this case the two sets of 
lines almost coincide, indicating that the numerical solution is close
to being force-free. Some discrepancy is observed for long field lines
because of the need to interpolate values of $\vec J$ and $\vec B$ 
between grid points in order to draw field lines. At each grid point $\vec J \times \vec B$
is very small (as discussed below), but larger values are obtained between
grid points when $\vec J$ and $\vec B$ are interpolated. The field lines
which show the largest discrepancy occur close to the boundary separating
zero and non-zero values of $\alpha_0$, where the interpolation is least
accurate.

\begin{figure}
  \centerline{\includegraphics[scale =0.75]{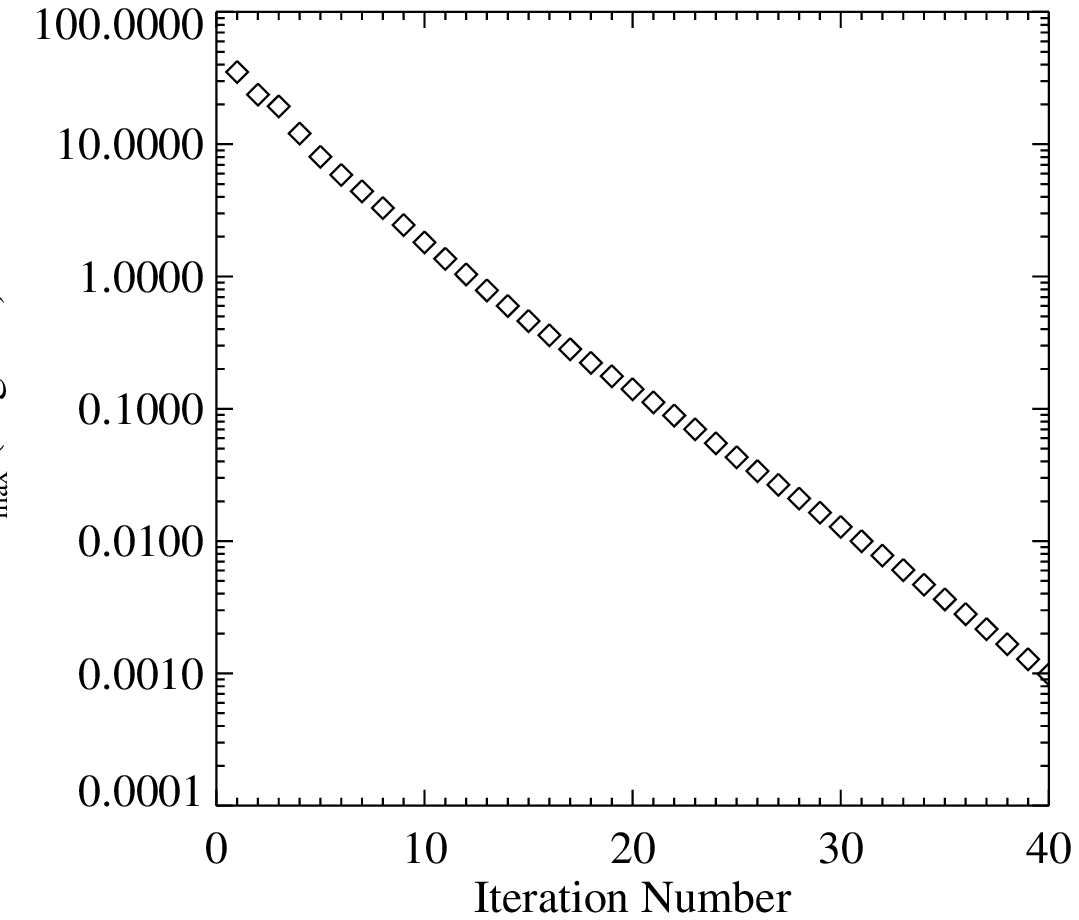}\includegraphics[scale =0.75]{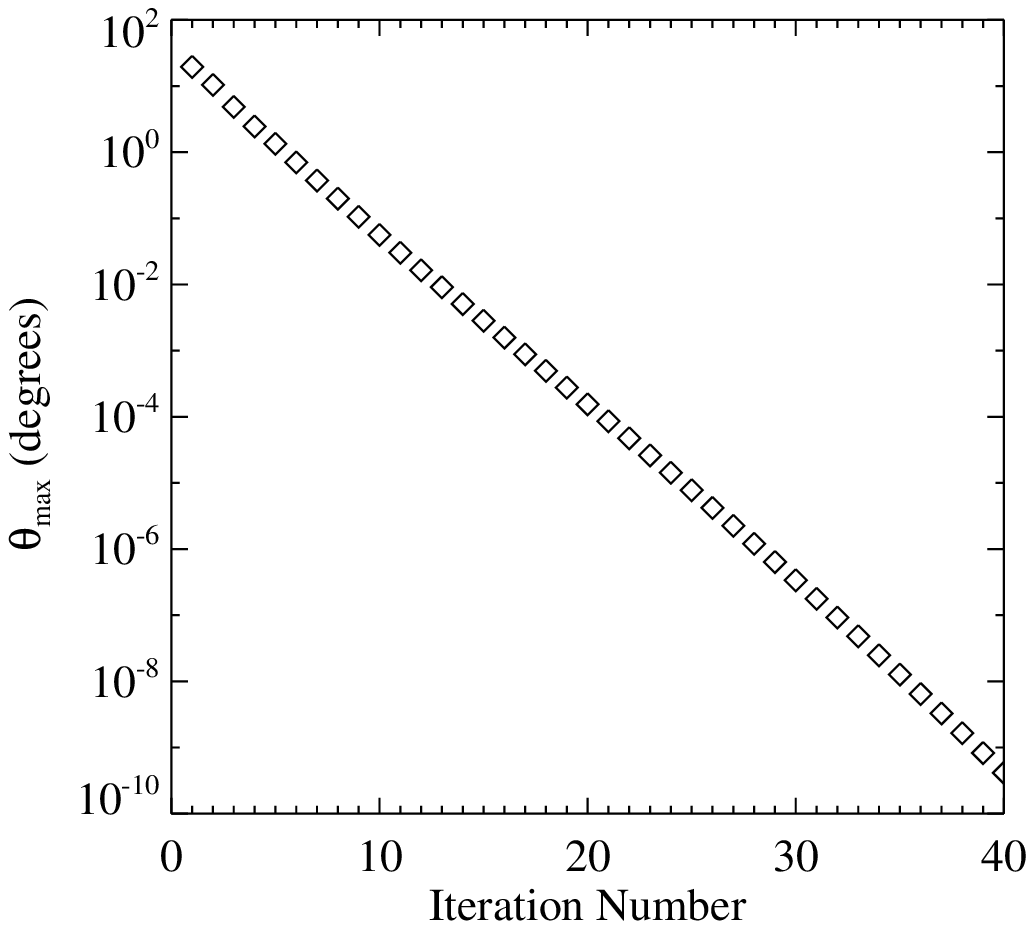}}
  \caption{The maximum angle between $\vec J$ and $\vec B$ over the solution domain as a function
           of iteration number. The left panel shows $\theta_{\rm max}$ 
           for the first test case in $\Omega_{\rm global}$, and the right panel 
           shows $\theta_{\rm max}$ for the second test case in $\Omega_{\rm wedge}$.
           The vertical scale is logarithmic. } 
\label{f4}
\end{figure}

The left panel of Figure $\ref{f4}$ shows $\theta_{\rm max}$ as a function 
of iteration number. After the first iteration the maximum angle between
$\vec J$ and $\vec B$ is approximately $20$ degrees, and $\theta_{\rm max}$
decreases approximately exponentially over the 40 iterations to a 
final value of order $10^{-3}$ degrees. This confirms that a force-free 
solution is found.

%
%

\subsection{Results for test case two: the bipole in the restricted domain}
\label{re_sec2}

\begin{figure}
  \centerline{\includegraphics[scale =0.75]{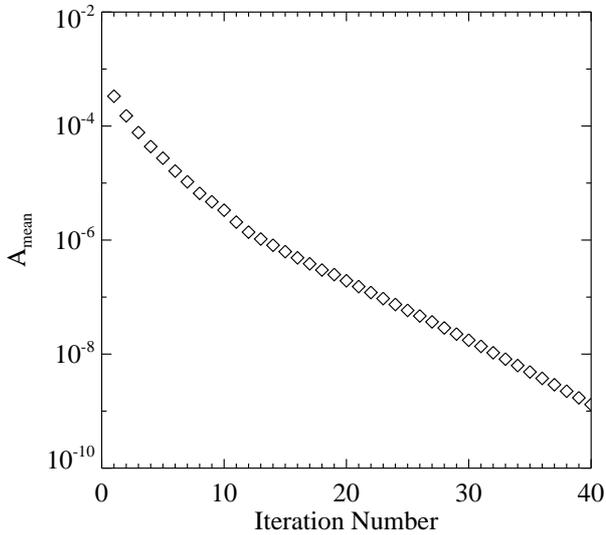}\includegraphics[scale =0.75]{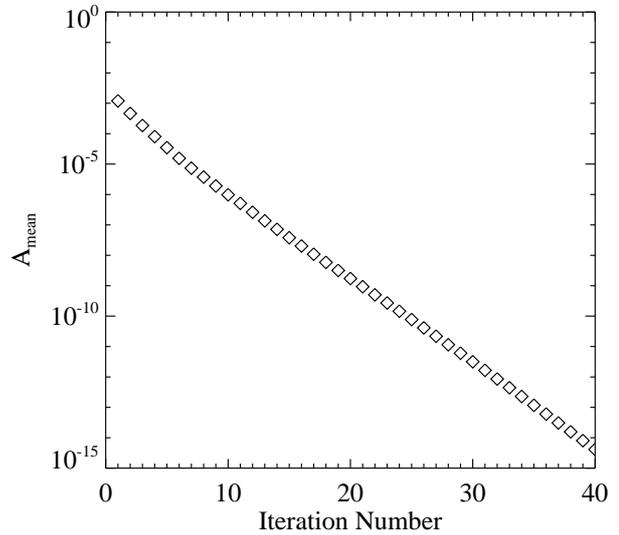}}
  \caption{The average absolute change in the field, $A_{\rm mean}$, 
   as a function of iteration number for the first test case (left panel)
   and the second test case (right panel). The vertical scale is 
   logarithmic.} 
\label{f2}
\end{figure}

The second test case described in Section \ref{sec_btc} considers the bipolar 
test case in the domain $\Omega_{\rm wedge}$.
We choose a domain which spans $20^{\circ}$ in latitude and $20^{\circ}$ in longitude. The size
and the dimensions of the numerical grid are summarized in Table \ref{t2}.
In this case, the separation between the spots is an order of magnitude smaller than
for the first test case, so we increase $\alpha_{\rm s}$ by 
an order of magnitude so that the product $L_{\rm s} \alpha_{\rm s}$ matches
test case one. This provides a bipole with the same
amount of twist. The parameters for this test
case are summarized in Table \ref{f2}. We again apply 40 Grad-Rubin iterations,
which takes 1.6 hours to run on a computer with an eight-core CPU.

The right panel of Figure \ref{f1} shows the energy (in units of the energy of the 
potential field $E_0$) as a function of iteration number. The energy converges to a constant value 
after approximately four Grad-Rubin iterations. The right panel of Figure \ref{f2} shows
$A_{\rm mean}$ as a function of iteration number. The scale on 
the $y$ axis is logarithmic, and the figure shows
that $A_{\rm mean}$ decreases approximately exponentially, to a final value 
the order of machine precision. The behavior of $E/E_0$ and 
$A_{\rm mean}$ as a function of iteration confirm  the convergence of the 
Grad-Rubin iteration procedure.

\begin{figure}
              \includegraphics[scale =0.5]{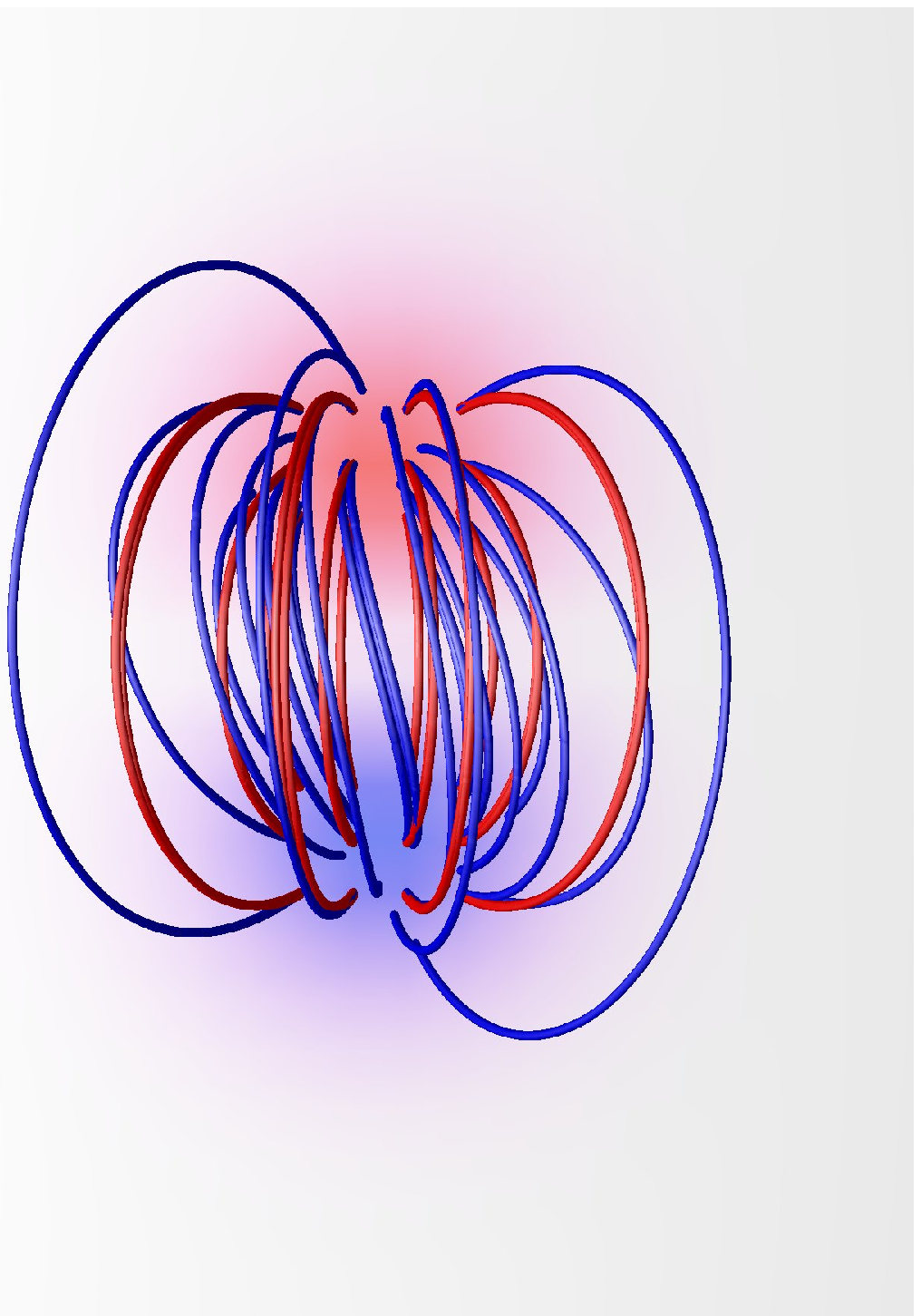}  
              \includegraphics[scale =0.5]{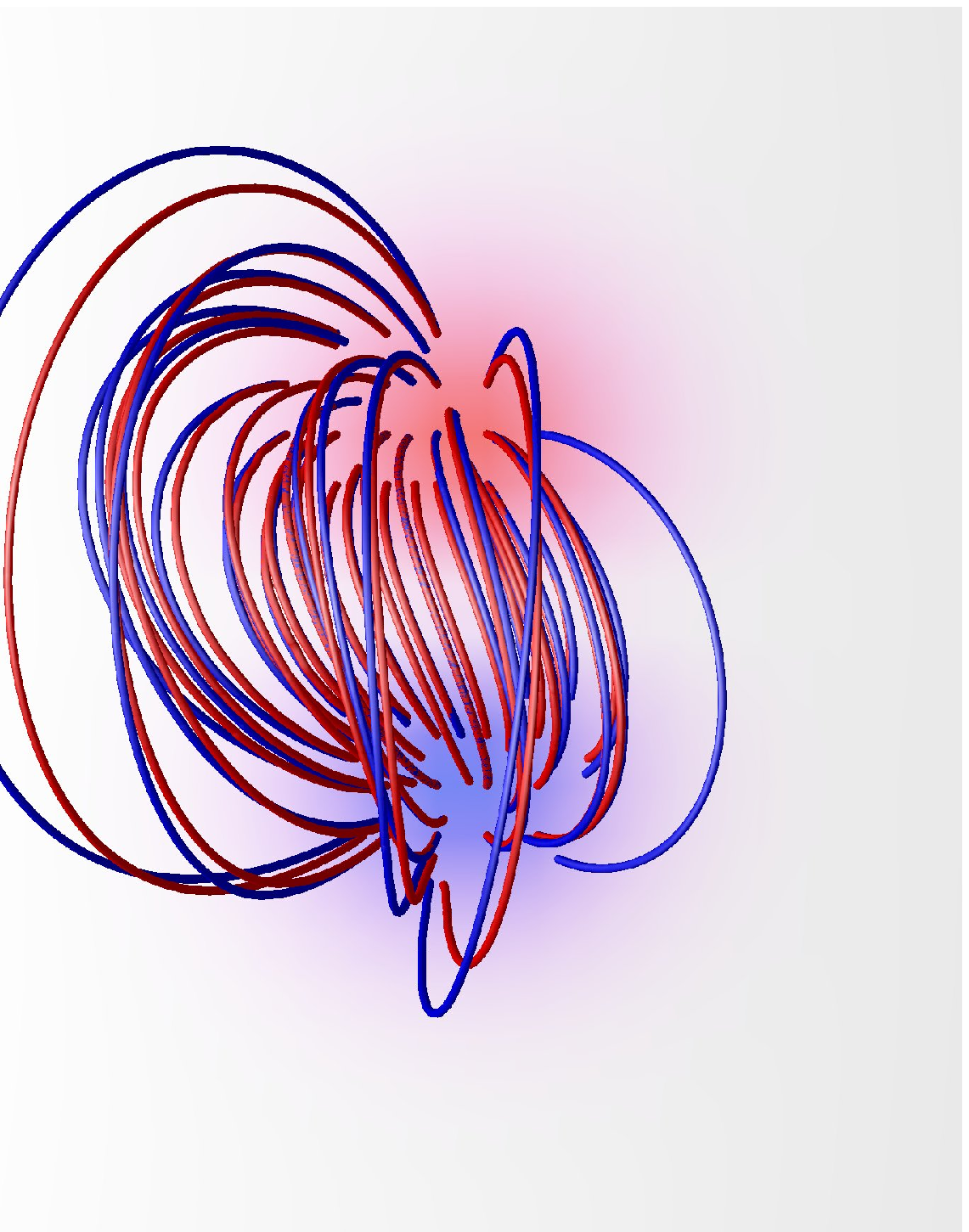}
  \caption{Field lines (blue) and streamlines of current density (red)
           for test case two. The left panel is after the first 
           Grad-Rubin iteration. The right is after 40 Grad-Rubin iterations. In
           the left panel there is a clear difference between the two 
           sets of lines, and in the right the two sets of lines coincide, indicating 
           that a force-free solution has been found. The discrepancy between the two 
           sets of lines is due to the interpolation of field values between
           grid points necessary to draw the lines (see the discussion in Section \ref{tcone_bg}).} 
\label{f5}
\end{figure}

The left panel of Figure \ref{f5} shows the field lines of the magnetic field
and the streamlines of the current density after the first iteration (left panel) and after the last iteration (right panel). 
The magnetic field is shown in blue and the current density in red. 
In the left panel there is a clear difference between the two sets of lines, 
and in the right panel the two sets of lines closely coincide, indicating that 
a force-free solution is found. Some disagreement is observed between the 
two sets of lines for longer field lines and is explained in Section \ref{tcone_bg}. 

The right panel of Figure $\ref{f4}$ shows $\theta_{\rm max}$ as a function 
of iteration number. After the first iteration the maximum angle between
$\vec J$ and $\vec B$ is approximately $20$ degrees, and $\theta_{\rm max}$
decreases approximately exponentially with iteration, to a final
value of order $10^{-9}$ degrees.

%
%

\section{Discussion and conclusion}
\label{sec5}

We present an implementation of the Grad-Rubin method \citep{gr} for 
solving the force-free boundary value problem in spherical polar
geometry with the aim of modeling the solar coronal magnetic field. The 
method is applicable to either the entire corona or a wedge with
restricted angular extent. 

We apply our code to two test cases with analytic boundary conditions,
representing a current-carrying bipole, to demonstrate the convergence 
of the Grad-Rubin iteration and the self-consistency of 
the final numerical solution. For the first case we compute a 
solution in a domain which encompasses the entire corona, and in 
the second we compute a solution in a spherical wedge. 
We find that in both cases the Grad-Rubin iteration converges, and
the resulting solution is force free to a very good approximation, by which we 
mean that the angle between $\vec J$ and $\vec B$ is small. 

The bipolar test case we consider is simple, but serves
to illustrate the basic correctness of the method. We are 
unable to identify a suitable exact analytic equilibria with
which to test the code. The \citet{1990ApJ...352..343L} field has often been
used for testing force-free codes \citep{1999A&A...350.1051A,2007SoPh..240..227W},
but it proves difficult to reproduce this solution without imposing 
analytic boundary conditions on an outer shell located at a finite 
radius from  the photosphere. Using the asymptotic boundary condition defined by Equation
(\ref{asym}), requires a very large domain to
encompass all the significant currents for the \citet{1990ApJ...352..343L} problem. 
Hence it is difficult to calculate accurate solutions without resorting
to the use of exact boundary conditions at a finite outer shell.
The lack of an exact analytic
test case means that we cannot determine how the numerical error 
scales as a function of grid resolution.
We also note that our test case has smooth boundary 
conditions. In practice, we expect observational data to contain significant
gradients, in particular due to noise which is present in the data. The 
effect of Gibbs phenomenon produced by the spectral method applied to these 
large gradients has not yet been investigated.

We note that the code takes 1.6 hours to run for the test case in
the spherical wedge, for a grid with $64^3$ points. By comparison the calculation
in the global domain takes 15 minutes for a $128 \times 64 \times 128$
grid. The difference in speed occurs because the spherical harmonics
must be summed to large orders when $\Delta \phi$ is small. In the 
first case $L=64$ and in the second case $L=567$. In practice it may not be necessary
to use an $L$ value of a given size for all values of $r$. We expect
that as the field becomes smoother with increasing $r$ higher order
terms may be ignored and a reduction in $L$ as a function of $r$ may
increase the speed of the code. Also, we expect that it may be possible
to increase the speed of the code by improving details of the technique used
for the calculation of the non-potential field. These improvements will be 
implemented in future work. 

As mentioned in Section \ref{ss_ssftpf}, \citet{2011ApJ...732..102T} 
report the non-convergence of the spherical harmonics spectral series for large $L$,
for the calculation of potential field models. The grid used
by \citet{2011ApJ...732..102T} is uniform in $\cos\theta$. However, near the
poles, the associated Legendre polynomials are rapidly varying functions
of $\cos\theta$, and the rapid variations are not accurately represented on
the given grid. This results in the non-convergence of the series. For our calculations
in $\Omega_{\rm wedge}$, we use a grid uniform in $\theta$. 
\citet{2011ApJ...732..102T} report that such a grid does not cause convergence problems. 
Furthermore, the region considered in Section \ref{re_sec2} is sufficiently isolated
from the poles that convergence problems are unlikely to occur regardless
of the grid used. When considering larger regions which cannot be isolated
from the poles, or for calculations in $\Omega_{\rm global}$ we use a 
Gauss-Legendre grid which is dense at the poles and accurately 
represents the rapid variations in the associated Legendre polynomials.

Other developments to the code are also planned. 
At present, the code uses a uniform grid in $r$, but this could
be changed to a nonuniform grid. A nonuniform grid should be more efficient because 
the grid can be chosen to be dense close to the photosphere where the magnetic field is 
structured on small scales, and sparse far from the photosphere where the field is smooth. 
At present the method assumes the asymptotic boundary condition
Equation (\ref{asym}) but this could be changed to accommodate
boundary conditions at an outer ``source surface" analogous to the 
potential source-surface model. Finally, the method 
presently only uses $\alpha_0$ over a single polarity, which is likely to be  
problematic when using observational data because of the inconsistency
of the data with the force-free model \citep{2008ApJ...675.1637S}. In 
future work we will modify our method to implement the self-consistency 
procedure of \citet{2009ApJ...700L..88W}.

The method and code outlined in this paper are designed for
 application to solar data, but we have not yet attempted
this. In future work we will test the code on vector 
magnetogram data derived from observations by the {\it Helioseismic
and Magnetic Imager} aboard the {\it Solar Dynamics Observatory} (SDO/HMI). 
The SDO/HMI instrument provides data for the whole solar disk, which demands spherical modeling.
We hope to be able to develop the code presented here into a practical tool for modeling of
the coronal magnetic field from SDO/HMI data, and for application to other, future data sets.

%
%
\section*{Appendix A}

In this appendix we derive Equations (\ref{b1_comp})-(\ref{b3_comp})
using vector spherical harmonics. 

For any magnetic field it is possible to introduce a vector 
potential, $\vec A$, related to the magnetic field by  
\BE
  \nabla \times \vec A = \vec B. 
\EE
In the Coulomb gauge, 
\BE
  \nabla \cdot \vec A = 0,
\EE
in which case the vector potential is given by 
the vector Poisson equation, \citep{1998clel.book.....J}:
\BE
  \nabla^2 \vec A = -\mu_0 \vec J.
  \label{poss}
\EE
The boundary conditions on $\vec A$
enforcing the boundary conditions on $\vec B$ given by Equations (\ref{np_bc1}) and (\ref{np_bc2}) 
are
\BE
  \vec A \times {\hat \vec r} = 0,
\EE
and 
\BE
   \lim_{r \rightarrow \infty} \vec A = 0
\EE
respectively. 

We can solve Poisson's equation by expanding $\vec A$ in terms of 
a set of orthonormal basis functions, which reduces the partial differential
equation to a system of ordinary differential equations for the 
series coefficients. The vector spherical harmonics provide a natural set of 
basis functions in spherical polar coordinates \citep{1953mtp..book.....M}. The vector 
potential can be written in terms of these functions as:
\BE
  \vec A = \sum \limits_{l=0}^{\infty} \sum \limits_{m=-l}^{l} 
  A^{(1)}_{lm} \vec Y_{lm} + A^{(2)}_{lm} \vec \Psi_{lm} + A^{(3)}_{lm} \vec \Phi_{lm},
  \label{apA_A1}
\EE
where $A^{(i)}_{lm}$ are the spectral coefficients with $i=1,2,3$. 
By substituting Equation (\ref{apA_A1}) into Equation (\ref{poss}) and using the orthonormality 
of the basis functions, it can be shown that $A^{(3)}_{lm}$ satisfies
the second-order linear inhomogeneous equation 
\BE
  \frac{d^2 A^{(3)}_{lm}}{dr^2}+\frac{2}{r}\frac{d A^{(3)}_{lm}}{dr}-\frac{l(l+1)}{r^2}A^{(3)}_{lm} 
  = -\mu_0 J_{lm}^{(3)},
  \label{A3_eq}
\EE
where
\BE
  J_{lm}^{(3)} = \int \limits_0^{\pi} \int \limits_0^{2\pi} \vec J \cdot \vec \Phi_{lm} \sin\theta d\phi d\theta.
\EE
Equation (\ref{A3_eq}) has a general solution related to the 
boundary conditions on $\vec A$, and a particular solution,  
determined by the source terms $J_{lm}^{(3)}$. The solutions can be 
found analytically using the method of variation of parameters \citep{1989hde..book.....Z}. 
This gives
\BE
  A^{(3)}_{lm}(r)=- R_{\sun}^l \left (\frac{R_{\sun}}{r} \right )^{l+1}I_0  + I_1(r) + I_2(r),
  \label{aa_vp3}
\EE
where 
\BE
  I_0 = \frac{\mu_0}{2l+1} \int^{\infty}_{R_{\sun}} s^{1-l} J^{(3)}_{lm}(s) ds,
  \label{aa_int0}
\EE
\BE
  I_1(r) = \frac{\mu_0}{2l+1} \int^{r}_{R_{\sun}} s \left ( \frac{s}{r} \right )^{l+1}J^{(3)}_{lm}(s)ds,
  \label{aa_int1}
\EE
and 
\BE
  I_2(r) = \frac{\mu_0}{2l+1} \int^{\infty}_{r} s \left ( \frac{r}{s} \right )^l J^{(3)}_{lm}(s)ds.
  \label{aa_int2}
\EE
The spectral coefficients $B^{(1)}_{lm}$ and $B^{(2)}_{lm}$ can be 
determined from Equation (\ref{aa_vp3}) as follows. Taking the
curl of the vector spherical harmonics leads to the following
identities 
\BE
  \nabla \times [F_{lm}(r) \vec Y_{lm} ] = \sqrt{l(l+1)} \frac{F_{lm}(r)}{r} \vec \Phi_{lm},
  \label{curl1}
\EE
\BE
  \nabla \times [F_{lm}(r) \vec \Psi_{lm} ] = -\left ( \frac{d}{dr} + \frac{1}{r} \right ) F_{lm}(r)\ \vec \Phi_{lm},
  \label{curl2}
\EE
and 
\BE
  \nabla \times [F_{lm}(r) \vec \Phi_{lm} ] = \sqrt{l(l+1)}\frac{F_{lm}(r)}{r}\vec Y_{lm}
  + \left ( \frac{d}{dr} + \frac{1}{r} \right ) F_{lm}(r)  \vec \Psi_{lm},
  \label{curl3}
\EE
where $F_{lm}(r)$ is a any function of $r$. 
Since $\vec B = \nabla \times \vec A$, it follows that
\begin{eqnarray}
  \vec B & = & \sum \limits_{l=0}^{\infty} \sum \limits_{m=-l}^{l} 
  \sqrt{l(l+1)}\frac{A^{(3)}_{lm}(r)}{r}\vec Y_{lm}
  + \left ( \frac{d}{dr} + \frac{1}{r} \right ) A^{(3)}_{lm}(r)  \vec \Psi_{lm} \\
  &+& \left [ \sqrt{l(l+1)} \frac{A_{lm}^{(1)}(r)}{r}  -
  \left ( \frac{d}{dr} + \frac{1}{r} \right ) A_{lm}^{(2)}(r) \right ] \vec \Phi_{lm}\nonumber.
  \label{b_curl}
\end{eqnarray}
The spectral coefficients are then given by 
\BE
  B^{(1)}_{lm} = \sqrt{l(l+1)}\frac{A^{(3)}_{lm}(r)}{r},  
  \label{aa_b1}
\EE
and
\BE
  B^{(2)}_{lm} =\left ( \frac{d}{dr} + \frac{1}{r} \right ) A^{(3)}_{lm}(r),
  \label{aa_b2}
\EE
where we use the fact that the vector spherical harmonics 
are orthonormal. By substituting Equation (\ref{aa_vp3}) into 
Equations (\ref{aa_b1}) and (\ref{aa_b2}) we find that 
\BE
  B^{(1)}_{lm} = \frac{\sqrt{l(l+1)}}{r} \left [ -R_{\sun}^l \left ( \frac{R_0}{r} \right )^{l+1} I_0 + I_2(r) + I_3(r)
  \right ],
\EE
and
\BE
  B^{(2)}_{lm} = \frac{1}{r} \left [ R_{\sun}^l \left (\frac{R_{\sun}}{r} \right )^{l+1}I_0
  - lI_1(r) + (l+1)I_2(r)  \right ].
\EE
It is possible to determine $B^{(3)}_{lm}$ by finding $A^{(1)}_{lm}$ and
$A^{(2)}_{lm}$ in the same manner as for $A^{(3)}_{lm}$, and
evaluating the expression in Equation (\ref{b_curl}) for the coefficients
of $\vec \Phi_{lm}$. However, a simpler approach is to apply Equation (\ref{curl3}) to
Ampere's law $\nabla \times \vec B = \mu_0 \vec J$. This 
yields 
\BE
  B^{(3)}_{lm} =  \frac{r J^{(1)}_{lm}}{\sqrt{l(l+1)}},
\EE
where
\BE
  J_{lm}^{(1)} = \int \limits_0^{\pi} \int \limits_0^{2\pi} \vec J \cdot \vec Y_{lm} \sin\theta d\phi d\theta.
\EE

%
%
\section*{Appendix B}

In this appendix we outline a parallel method for evaluating
the sums in Equations (\ref{br_pot})-(\ref{bp_pot}). 
The method of summing the spectral series is important because it 
determines the speed of the computation. Also, a high resolution grid requires
a large order $L$, in which case the memory required to compute the
spectral solution becomes large (depending of the implementation).  
In the following we present a method for summing the series which 
is fast and which uses memory efficiently. The procedure is described for 
the $B_r$ component of the potential field. A similar approach is 
used for the other components of the potential field, and for the 
non-potential field. 

\subsection{Parallel summation of the spectral series}

The following procedure for computing $B_r$ follows a prescription
in \citet{2007nrfa.book.....P}. Here we briefly outline the method  
and describe a parallel implementation. 

The spectral series for $B_r$ can be written as 
\BE
B_r = \sum \limits_{m=-L}^{L} g_m(r,\theta) e^{im\phi},
  \label{fft1}
\EE
where the auxiliary function $g_m(r,\theta)$ is defined as 
\BE
  g_m(r,\theta) = \sum \limits_{l=|m|}^{L} a_{lm} \left (\frac{R_{\sun}}{r} \right)^{l+2} 
  \tilde{P}^m_l(\cos\theta),
  \label{aux1}
\EE
and where the spherical harmonics $Y_{lm}$ are written as a product of a 
normalized-associated Legendre polynomial ${\tilde P}^m_l(x)$ and 
the complex exponential. In our method we first compute 
the auxiliary function using Equation (\ref{aux1}), and then 
determine $B_r$ using Equation (\ref{fft1}).

This procedure can be implemented in parallel by partitioning the sum in Equation 
(\ref{fft1}) into $N$ partial sums which are computed 
independently. We define the partial sum  
\BE
F_n = \sum \limits_{m=-m_n}^{m_n} g_m(r,\theta) e^{im\phi},
\EE
where 
\BE
m_n = nL/N.
\EE
The complete solution is then  
\BE
  B_r = \sum \limits_{n=0}^L F_n. 
\EE
Each $F_n$ can be computed independently, allowing trivial parallelism.
Most high performance computer clusters consist of a series of nodes, with 
each node containing several processors with shared a memory space. 
Our code distributes the computation of $F_n$ among such nodes using the 
Message Passing Interface (MPI) \citep{mpiref}. The computation of $F_n$ 
at each node is parallelized across the processors on the node 
using OpenMP \citep{openmp}. 

\subsection{Memory requirements}

It is important to consider the memory required to store $g_m(r,\theta)$ as an array. 
Since the integrand in Equation (\ref{co1}) is real, the coefficients with
$m<0$ contain the same information as those with $m >0$ \citep{2007nrfa.book.....P}. 
Therefore, it is necessary only to store coefficients with $m \ge 0$, 
and there are $(L+1)L/2$ such coefficients. 
Therefore an array storing $g_m(r,\theta)$ has $L(L+1) N_r N_{\theta}/2$ 
elements for a grid of size $N_r \times N_{\theta} \times N_{\phi}$.
The associated memory use may prove problematic for $L\sim 1000$. A grid
of size $N_r=128$, $N_{\theta} = 256$, and $N_{\phi} = 512$ requires
$\approx 100 \, {\rm Mb}$ to store $B_r$ using double precision floating 
point numbers, and $\approx 1 \, {\rm Gb}$ to store $g_m(r,\theta)$.
The memory requirements for $g_m(r,\theta)$ can quickly become
larger than a few gigabytes, a typical size for total memory on
current desktop computer hardware.

The partial summation method reduces the memory requirements because
the array $g_m$ is split among the nodes. If the series is split into
$N$ partial sums, then the memory requirement for each node is $\sim 1/N$ 
of that needed to store the entire array. The partial summation method
can also be used to conserve memory with only a single node. In this
case each partial sum is computed sequentially. This way only part 
of the spectrum $g_m$ is computed for each partial summation. Again,
this method reduces the memory requirements by a factor $\sim 1/N$
compared with a single serial summation.


%
%
\begin{table}
\caption{Parameters used for the bipolar test cases involving the domain
$\Omega_{\rm global}$ (case 1), and the wedge-shaped region $\Omega_{\rm wedge}$
(case 2). Distances are quoted in units of a solar radius, and magnetic field strengths are relative
to the scale constant $B_{\rm s}$. Units are given in square brackets. }
\begin{tabular}{l l l  l  l l l }
\hline
\hline             
Case  & $\sigma$ & $ L_{\rm s}$    & $\Lambda$ & $\alpha_{\rm s} L_{\rm s}$  & $B_{\rm th}$\\   
      & [$R_{\sun}$]  & [$R_{\sun}$]         & [$R_{\sun}^{-1}$] &                         & [$B_s$]     \\        
\hline
1    & $0.1$   &  $0.35$ &  3  &   1.05       & 0.75 \\ 
2    & $0.02$   &  $0.035$ &  30 &    1.05         & 0.75 \\      
\end{tabular}
\label{t1}
\end{table}

%
%
\begin{table}
\caption{Size of the grids used for test cases. Both the physical
sizes and the sizes of the numerical grids are shown.}
\begin{tabular}{l l l  l  l l l l}
\hline
\hline             
Case & Domain & $N_{r}$ & $N_{\theta}$    & $N_{\phi}$ & Lon & Lat & $R_{\rm max}$ \\   
    &         &   & &                                  & [degrees] & [degrees] & [$R_{\sun}$] \\   
\hline
1    & $\Omega_{\rm global}$ & 128   &  64 &  128  &   360      & 180  & 6 \\ 
2    & $\Omega_{\rm wedge}$ & 64   & 64 &  64 &       20         & 20 & 1.2 \\      
\end{tabular}
\label{t2}
\end{table}

%
%

\begin{acks}
  S. A. Gilchrist acknowledges the support of an Australian 
  Postgraduate Research Award. 
\end{acks}


\end{article} 

\begin{thebibliography}{}
%
\bibitem[Alissandrakis(1981)]{1981A&A...100..197A} 
  Alissandrakis, C.E.: 1981, {\it Astron. Astrophys.} {\bf 100}, 197.
%
\bibitem[Altschuler and Newkirk(1969)]{1969SoPh....9..131A}  
  Altschuler, M.D., Newkirk, G.: 1969, {\it Solar Phys.} {\bf 9}, 131. doi:10.1007/BF00145734.
%
\bibitem[Amari, Boulmezaoud, and Mikic(1999)]{1999A&A...350.1051A} 
  Amari, T., Boulmezaoud, T.Z., Mikic, Z.: 1999, {\it Astron. Astrophys.} 
  {\bf 350}, 1051. 
%
\bibitem[Amari et al.(2013)]{2013A&A...553A..43A} 
  Amari, T., Aly, J.J., Canou, A., Mikic, Z.: 2013, {\it Astron. Astrophys.} 
  {\bf 553}, A43. doi:10.1051/0004-6361/201220787.
%
\bibitem[Barrera, Est{\'e}vez, and Giraldo,(1985)]{vshmag} 
  Barrera, R.G., Est{\'e}vez, G.A., Giraldo, J.: 1985, {\it Eur. J. Phys.}
  {\bf 6}, 287.
%
\bibitem[Boyd(2001)]{boyd_spectral} 
  Boyd, J.P.: 2001, {\it Chebyshev and Fourier Spectral Methods}, 2nd
  edn. Dover Publications Inc, New York, 380. 
%
\bibitem[Chandra \emph{et al.}(2001)]{openmp} 
  Chandra, R., Menon, R., Dagum, L., Kohr, D., Maydan, D., 
  McDonald, J.: 2001, {\it Parallel Programming in OpenMP},
  Morgan Kaufmann Publishers, San Francisco, 1. 
%
\bibitem[De Rosa \emph{et al.}(2009)]{2009ApJ...696.1780D} 
  De Rosa, M.L., Schrijver, C.J., Barnes, G., Leka, K.D., Lites, B.W., 
  Aschwanden, M.J., et al.: 2009, {\it Astrophys. J.} {\bf 696}, 1780.
  doi:10.1088/0004-637X/696/2/1780.
%
\bibitem[Dennis and Quartapelle(1985)]{denquart} 
  Dennis, S.R., Quartapelle, L.: 1985, {\it J. Comp. Phys.}
  {\bf 61}, 218.
%
\bibitem[Gary(2001)]{2001SoPh..203...71G} 
  Gary, G.A.: 2001, {\it Solar Phys.} {\bf 203}, 71. 
%
\bibitem[Gary and Hagyard(1990)]{1990SoPh..126...21G} 
  Gary, G.A., Hagyard, M.J.: 1990, {\it Solar Phys.} {\bf 126}, 21. 
%
\bibitem[Grad and Rubin(1958)]{gr} 
  Grad, H., Rubin, H.: 1958, {\it Proc. 2nd Int. Conf. Peaceful 
  Uses of Atomic Energy,} {\bf 31}, 190.
%
\bibitem[Jackson(1998)]{1998clel.book.....J} 
  Jackson, J.D.: 1998, {\it Classical Electrodynamics,} 
  3rd edn. Wiley, New York, 180.  
%
\bibitem[Landi Degl'Innocenti and Landolfi(2004)]{2004ASSL..307.....L} 
  Landi Degl'Innocenti, E., Landolfi, M.: 2004,
  {\it Polarization in Spectral Lines}, Kluwer Academic Publishers, Dordrecht,
  625. 
%
\bibitem[Low and Lou(1990)]{1990ApJ...352..343L} 
  Low, B.C., Lou, Y.Q.: 1990, {\it Astrophys. J.} {\bf 352}, 343. doi:10.1086/168541.
%
\bibitem[Metcalf \emph{et al.}(1995)]{1995ApJ...439..474M} 
  Metcalf, T.R., Jiao, L., McClymont, A.N., Canfield, R.C., Uitenbroek, H.: 1995, {\it 
  Astrophys. J.} {\bf 439}, 474. doi:10.1086/175188.
%
\bibitem[Molodensky(1974)]{1974SoPh...39..393M} 
  Molodensky, M.M.: 1974, {\it Solar Phys.} {\bf 39}, 393. doi:10.1007/BF00162432.
%
\bibitem[Morse and Feshbach(1953)]{1953mtp..book.....M} 
  Morse, P.M., Feshbach, H.: 1953, {\it Methods of Theoretical Physics, 
  Part Two}, McGraw-Hill, New York, 1252.  
%
\bibitem[Nakagawa and Raadu(1972)]{1972SoPh...25..127N} 
  Nakagawa, Y., Raadu, M. A.: 1972,  {\it Solar Phys.} {\bf 25}, 127. doi:10.1007/BF00155751. 
%
\bibitem[Pesnell, Thompson, and Chamberlin,(2012)]{2012SoPh..275....3P} 
  Pesnell, W.D., Thompson, B.J., Chamberlin, P.C.: 2012, {\it Solar Phys.}
  {\bf 275}, 3. doi:10.1007/s11207-011-9841-3.
%
\bibitem[Press \emph{et al.}(2007)]{2007nrfa.book.....P} 
 Press, W.H., Teukolsky, S.A., Vetterling, W.T., Flannery, B.P.: 2007,
 {\it Numerical Recipes. The Art of Scientific Computing},
 3rd edn. Cambridge University Press, Cambridge, 292. 
%
\bibitem[R{\'e}gnier(2009)]{2009A&A...497L..17R}
  R{\'e}gnier, S.\ 2009, {\it Astron. Astrophys.} {\bf 497}, 17. doi:10.1051/0004-6361/200811502. 
%
\bibitem[R{\'e}gnier(2012)]{2012SoPh..277..131R} 
  R{\'e}gnier, S.: 2012, {\it Solar Phys.} {\bf 277}, 131. doi:10.1007/s11207-011-9830-6.
%
\bibitem[Sakurai(1981)]{1981SoPh...69..343S} 
  Sakurai, T.: 1981, {\it Solar Phys.} {\bf 69}, 343. doi:10.1007/BF00149999.
%
\bibitem[Sakurai(1989)]{1989SSRv...51...11S} 
 Sakurai, T.: 1989, {\it Space Sci. Rev.} {\bf 51}, 11. doi:10.1007/BF00226267.
%
\bibitem[Schou et al.(2012)]{2012SoPh..275..229S} 
  Schou, J., Scherrer, P.H., Bush, R.I., Wachter, R., 
  Couvidat, S., Rabello-Soares, M.C., et al.: 2012, {\it Solar Phys.} {\bf 275}, 229. 
  doi:10.1007/s11207-011-9842-2.
%
\bibitem[Schrijver et al.(2008)]{2008ApJ...675.1637S} 
  Schrijver, C.J., De Rosa, M.L., Metcalf, T.R., Barnes, G.,
  Lites, B., Tarbell, T., et al.: 2008, 
  {\it Astrophys. J.} {\bf 675}, 1637. doi:10.1086/527413.
%
\bibitem[Snir \emph{et al.}(1998)]{mpiref} 
  Snir, M., Otto, S., Huss-Lederman, S., Walker, D., Dongarra., J: 1998,
  {\it MPI--the complete reference}, 2nd edn.
  MIT Press, Massachusetts, 1.
%
\bibitem[Sturrock(1994)]{1994ppit.book.....S}
  Sturrock, P.A.: 1994,
  {\it Plasma Physics, An Introduction to the Theory of Astrophysical,
  Geophysical and Laboratory Plasmas}, Cambridge University Press, Cambridge, 206.
%
\bibitem[Sun \emph{et al.}(2011)]{2011SoPh..270....9S} 
  Sun, X., Liu, Y., Hoeksema, J.T., Hayashi, K., Zhao, X.: 2011, {\it Solar Phys.}
  {\bf 270}, 9. doi:10.1007/s11207-011-9751-4.
%
\bibitem[T{\'o}th, van der Holst, and Huang(2011)]{2011ApJ...732..102T} 
  T{\'o}th, G., van der Holst, B., Huang, Z.: 2011,  {\it Astrophys. J.} {\bf 732}, 102. 
  doi:10.1088/0004-637X/732/2/102.
%
\bibitem[Valori, Kliem, and Keppens(2005)]{2005A&A...433..335V} 
  Valori, G., Kliem, B., Keppens, R.: 2005, {\it Astron. Astrophys.} {\bf 433}, 335. 
  doi:10.1051/0004-6361:20042008.
%
\bibitem[Wheatland(2007)]{2007SoPh..245..251W} 
  Wheatland, M.S.: 2007, {\it Solar Phys.} {\bf 245}, 251. 
  doi:10.1007/s11207-007-9054-y.
%
\bibitem[Wheatland and R{\'e}gnier(2009)]{2009ApJ...700L..88W} 
  Wheatland, M.S., R{\'e}gnier, S.: 2009, {\it Astrophys. J. Lett.} {\bf 700}, L88
  doi:10.1088/0004-637X/700/2/L88. 
%
\bibitem[Wheatland, Sturrock, and Roumeliotis(2000)]{2000ApJ...540.1150W} 
  Wheatland, M.S., Sturrock, P.A., Roumeliotis, G.: 2000, {\it Astrophys. J.}
  {\bf 540}, 1150. doi:10.1086/309355.
%
\bibitem[Wiegelmann(2004)]{2004SoPh..219...87W} 
  Wiegelmann, T.: 2004, {\it Solar Phys.} {\bf 219}, 87. doi:10.1023/B:SOLA.0000021799.39465.36.
%
\bibitem[Wiegelmann(2007)]{2007SoPh..240..227W} 
  Wiegelmann, T.: 2007, {\it Solar Phys.} {\bf 240}, 227. doi:10.1007/s11207-006-0266-3.
%
\bibitem[Wiegelmann(2008)]{2008JGRA..11303S02W} 
  Wiegelmann, T.: 2008, {\it J. Geophys. Res.} {\bf 113}, 3. doi:10.1029/2007JA012432.
%
\bibitem[Wiegelmann and Sakurai(2012)]{livrev2012} 
  Wiegelmann, T., Sakurai, T.: 2012, {\it Living Rev. Solar Phys.} {\bf 9}, 5. 
  \url{http://www.livingreviews.org/lrsp-2012-5}
%
\bibitem[Wu et al.(1990)]{1990ApJ...362..698W} 
  Wu, S.T., Sun, M.T., Chang, H.M., Hagyard, M.J., Gary, G.A.: 1990, 
  {\it Astrophys. J.} {\bf 362}, 698. doi:10.1086/169307.
%
\bibitem[Zwillinger(1989)]{1989hde..book.....Z} 
  Zwillinger, D.: 1989, {\it Handbook of Differential Equations}, 2nd
  edn. Academic Press, San Diego, 356. 
%
\end{thebibliography}
\end{document}